# The Morse lens

Huanyang Chen [1, 2,*]

[1] *Institute of Electromagnetics and Acoustics and Key Laboratory of Electromagnetic Science and Detection Technology, Xiamen University, Xiamen 361005, China*

[2] *Department of Electrical and Electronics Engineering, Xiamen University Malaysia, 43900 Sepang, Selangor, Malaysia*

**Abstract:** In this letter, we find that Morse potential (proposed about 90 years ago) could be connected to Coulomb potential (or Newton potential) and harmonic potential (or Hooke potential) by conformal mappings. We thereby design a new conformal lens from Morse potential, Eaton lens and Luneburg lens and propose a series of generalized Eaton/Luneburg lenses. We find that this Morse lens is a perfect self-focusing asymmetric lens, which differs from Mikaelian lens. Our theory provides a new insight to Morse potential and other traditional potentials and revisits their classical applications on designing lenses.

Traditional potentials, such as Coulomb potential and harmonic potential have been widely used in quantum mechanics. In addition, by analogy their classical potentials, e.g., Newton potential and Hooke potential, to geometric optics [1], people proposed Eaton lens [2] and Luneburg lens [3]. This kind of lenses are called absolute instruments and can perform a perfect imaging without aberration [4, 5] and have cavity modes with eigenfrequency spectra of tight groups that are almost equidistantly spaced [6]. Such properties in wave optics are helpful in designing conformal cloaks [7]. There are also many other absolute instruments, such as Maxwell's fisheye lens [8], Lissajous lens [9] and invisible lens [1]. In particular, recently people have found that Maxwell's fisheye lens could be connected to Mikaelian lens from an exponential conformal mapping [1, 10, 11, 12]. This lens is a symmetric one and has been implemented in visible frequencies from a curved waveguide and demonstrates the self-focusing and Talbot effect [11].

In fact, there are many other traditional potentials, such as Yukawa potential and Lennard-Jones (LJ) potential, which might also be useful in classical optical designs. In this letter, we will recall Morse potential proposed about 90 years ago [13]. We will prove that, from a conformal mapping aspect, it is actually related to Coulomb potential or harmonic potential. Therefore, we can propose a core conformal lens from this potential and generate Eaton lens, Luneburg lens, and even a more generalized form. We shall call this lens as Morse lens and provide the insight with the help of ray tracing simulations. We will also provide a new potential which we call the 6-10 potential for perfect imaging, and is different from the famous 6-12 potential or LJ potential. We hope to show a classical optics insight for this old potential and new lens and demonstrate the ability of lensing designs by combining quantum mechanics and transformation optics [7, 14, 15].

Let us start from Eaton lens with a profile of,

---

* kenyon@xmu.edu.cn

$$n = \sqrt{\frac{2}{r} - 1} \quad (1)$$

as shown by the red curve in Fig. 1a. By the connect of geometric optic and classical mechanics [1], i.e.,

$$n^2 = 2E - 2V \quad (2)$$

This lens is related to $V = -\frac{1}{r}$ for $E = -\frac{1}{2}$, which is a Coulomb potential or Newton potential. By performing an exponential conformal mapping [12], we can get another lens with a profile of,

$$n = \sqrt{2e^x - e^{2x}} \quad (3)$$

as shown by the red curve in Fig. 1b. Let us recall the Morse potential [13]

$$V = D(1 - e^{-a(x-x_0)})^2 - D \quad (4)$$

which is an ideal and typical anharmonic potential with exact solutions for diatomic molecules in quantum mechanics, where $D$ is the potential well depth, $a$ is the potential well width, and $x_0$ is the equilibrium position (which we will set as 0 in this letter). In addition, even in classical mechanics, this potential has analytical solutions [16]. By choosing $E = 0$, the conformal lens in Eq. (3) from Eaton lens is related to Morse potential with $D = \frac{1}{2}$ and $a = -1$.

Now we come to Luneburg lens with a profile of,

$$n = \sqrt{2 - r^2} \quad (5)$$

as shown by the green curve in Fig. 1a. This lens is related to $V = \frac{1}{2}r^2$ for $E = 1$, which is a harmonic potential or Hooke potential. With the same exponential mapping, the conformal lens with a profile of,

$$n = \sqrt{2e^{2x} - e^{4x}} \quad (3)$$

is obtained as shown by the green curve in Fig. 1b. This is also related to Morse potential with $D = \frac{1}{2}$ and $a = -2$ for $E = 0$.

We therefore propose a general lens with a profile of,

$$n = \sqrt{2e^{-ax} - e^{-2ax}} \quad (3)$$

which we shall call the Morse lens. And the related conformal lens with cylindrical or spherical symmetry could be written as,

$$n = \sqrt{2\frac{1}{r^{a+2}} - \frac{1}{r^{2a+2}}} \quad (4)$$

which we shall call the generalized Eaton/Luneburg lens and is related to a general LJ potential with $V = \frac{1}{2}(\frac{1}{r^{2a+2}} - \frac{2}{r^{a+2}})$ for $E = 0$. For example, for $a = 1$, $n = \sqrt{\frac{2}{r^3} - \frac{1}{r^4}}$, which is another form of Eaton lens with a repulsive potential core, and we shall call it an anti-Eaton lens. The refractive index profile is plotted by the blue curve in Fig. 1a and the related Morse lens is plotted by the blue curve in Fig. 1b, which is actually the same to that of Eaton lens by taking $x = -x$.

Now we perform ray tracing simulations for the above various lenses using COMSOL. For example, in Fig. 1c for Eaton lens, all rays from a point (say (1,0) here), will come back for self-imaging in elliptic trajectories. All the ellipses share the same focus point, i.e., the origin. While for the related Morse lens in Fig. 1d, all rays from a point (e.g., (-1, -16)) will have perfect imaging along x=-1 with a period of $2\pi$. For Luneburg lens in Fig. 1e, all rays from a point ((1,0) here) will have a perfect image at another point at the opposite side (i.e., (-1, 0) here). The trajectories are also ellipses but share the same center, the origin. The related Morse lens is plotted in Fig. 1f. All rays from a point ((-1, -16) here) will have perfect imaging along x=-1 with a period of $\pi$. Therefore, the intrinsic perfect imaging or self-imaging effect for Luneburg lens and Eaton lens is inherited from that of the Morse lens, with different periods. In fact, Eaton lens and Luneburg lens were proved to be connected simply by a power conformal mapping [1].

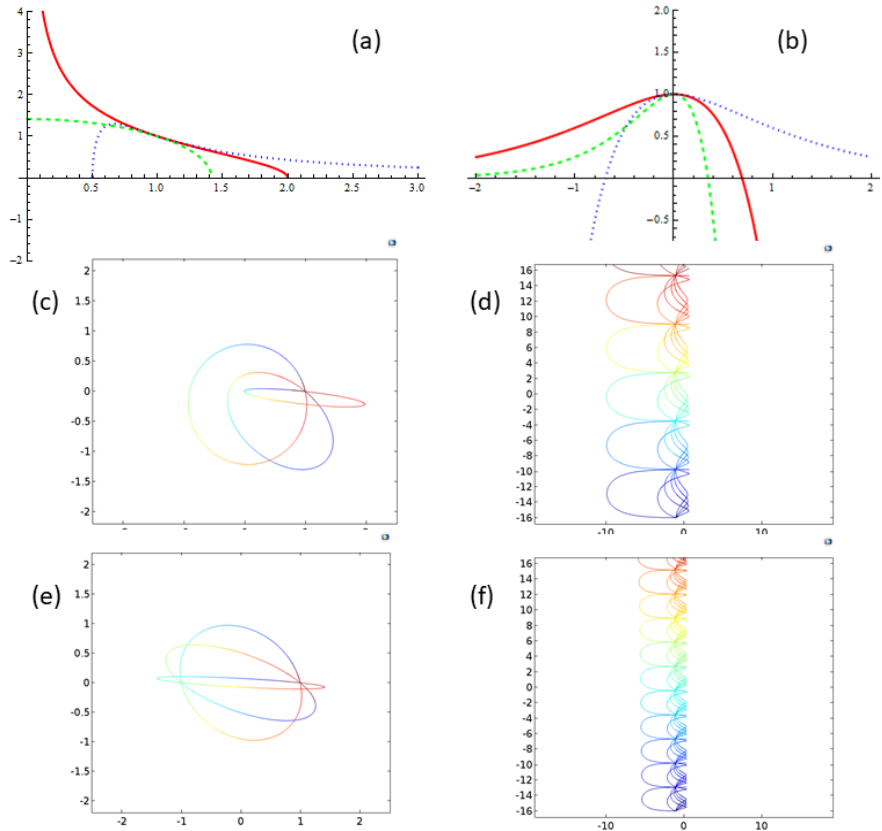

Fig. 1 (a) The refractive index profiles for Eaton lens (red), Luneburg Lens (green) and the anti-Eaton lens (blue). (b) The refractive index profile for the Morse lens with $a = -1$ (red), $a=-2$ (green) and $a=1$ (blue). (c) The ray trajectories from a point of (1, 0) in Eaton lens. (d) The ray trajectories from a point of (-1, -16) in Morse lens with $a=-1$. (e) The ray trajectories from a point of (1, 0) in Luneburg lens. (d) The ray trajectories from a point of (-1, -16) in Morse lens with $a=-2$.

Now we come to explore the effect of lenses in the generalized form. Similar to generalized Maxwell's fish-eye lens, for different order $a$, the trajectories are also different. In Fig. 2a-d, we plot the trajectory for each lens with (a) $a=-1$; (b) $a=-2$; (c) $a=-3$;(d) $a=-4$. We find that for $a=-3$, the trajectory is in a triangular shape, while for $a=-4$, the trajectory is in a square shape (for $a=-1$ and $a=-2$, they are ellipses). In addition, there are $abs(a)$ images (including self-image) for each lens (not plotted here). For a positive $a$, the central potential changes from attractive to repulsive. Therefore the trajectory is changed accordingly with $a$ concaves, as plotted in Fig. 2 for (e) $a=1$;(f) $a=2$;(g) $a=3$;(h) $a=4$. The number of images (including self-image) are $a$, which is similar to the attractive ones (we will only study one of them in the later section).

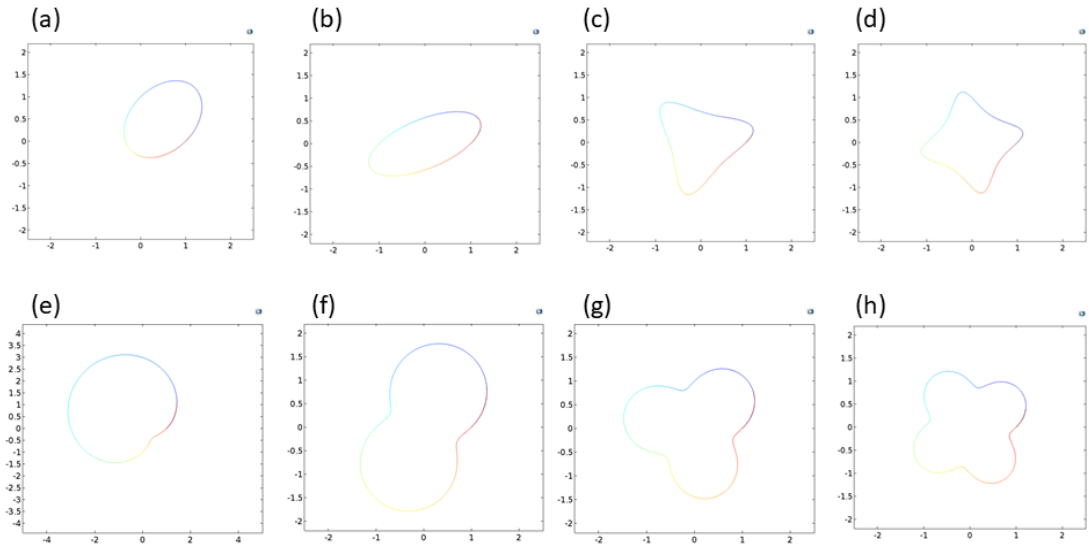

Fig. 2 The trajectory for a ray emitted from (1, 0) at $45^o$ in a generalized Eaton/Luneburg lens with (a) $a=-1$; (b) $a=-2$; (c) $a=-3$;(d) $a=-4$;(e) $a=1$;(f) $a=2$;(g) $a=3$;(h) $a=4$.

If $a=0$, the refractive index profile changes into that of an electromagnetic black hole [17, 18, 19,] i.e., $n=\dfrac{1}{r}$. If $a$ is not an integer but a rational number, we can always write it as $a=\mp\dfrac{p}{q}$.

For example, in Fig. 3a-c, we plot the trajectory for each lens with (a) $a=-\dfrac{1}{2}$; (b) $a=-\dfrac{3}{2}$; (c)

$a = -\frac{5}{2}$. There are $p$ images (including self-image) for each lens (not plotted here), while the light trajectories will travel around the origin $q$ times ($q\pi$) before it comes back as a self-image. For a positive $a$, the imaging properties are similar to the negative cases, while there are also $p$ concaves due to the repulsive potentials, see in Fig. 3 for (d) $a = \frac{1}{2}$;(e) $a = \frac{3}{2}$;(f) $a = \frac{5}{2}$. Note that the concave in Fig. 3d is very weak. After enlarging the figure, it would be clear. If $a$ is an irrational number, the lens is no longer an absolute instrument and there is no imaging or self-imaging effect.

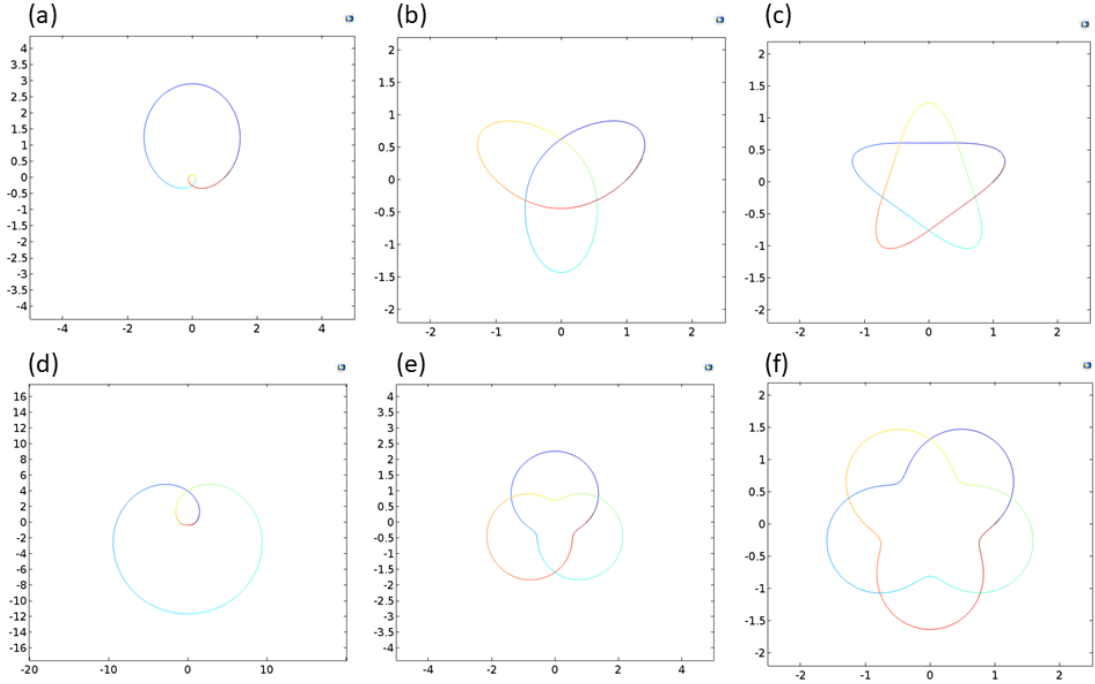

Fig. 3 The trajectory for a ray emitted from (1, 0) at 45° in a generalized Eaton/Luneburg lens with (a) $a = -\frac{1}{2}$; (b) $a = -\frac{3}{2}$; (c) $a = -\frac{5}{2}$;(d) $a = \frac{1}{2}$;(e) $a = \frac{3}{2}$;(f) $a = \frac{5}{2}$.

Finally, let us come to discuss why it is important to design a perfect imaging lens. Following our previous results [7, 20], any kind of cavity modes could be used in conformal cloaks. However, if the perfect imaging is valid, the cloaking effect will be very robust [7]. While for a general cavity mode, the effect would be compromised [20]. Moreover, it will also become useful to design devices to transfer information accurately, for example, the Mikaelian lens with Talbot effect. Therefore, to find more exact classical mechanics solutions similar to simple harmonic oscillators is vital [16]. In general, most of traditional potentials do not have this property. For example, the famous 6-12 potential or LJ potential (we will use $V = \frac{1}{2}(\frac{1}{r^{12}} - \frac{2}{r^6})$ to illustrate it) will not produce an absolute instrument. Nevertheless, by choosing $a = 4$, we will have a 6-10 potential ($V = \frac{1}{2}(\frac{1}{r^{10}} - \frac{2}{r^6})$), which is related to a generalized Eaton/Luneburg lens. Figure 4a shows the comparison between these two potentials. The red one is for the 6-10 potential, while the green one is for the 6-12 potential. Accordingly, we can obtain the related refractive index profiles in Fig. 4b. Although the potentials or refractive index profiles are very close to each other, the trajectories and imaging properties are very different. For the generalized lens, there are four perfect images (or self-image), as shown in Fig. 4c, which has also been discussed in the previous section. While for the related lens from 6-12 potential (whose refractive index profile is

$n = \sqrt{\dfrac{2}{r^6} - \dfrac{1}{r^{12}}}$ ), there is no perfect imaging effect, as shown in Fig. 4d.

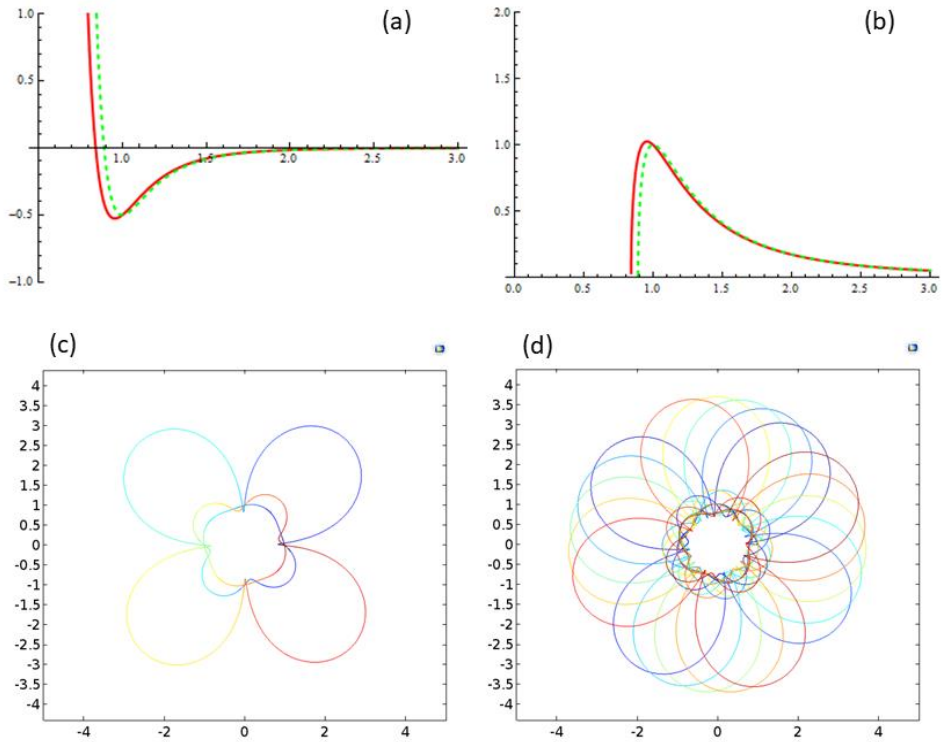

*Fig. 4 (a) The 6-10 potential (red) and the 6-12 potential (green). (b) The generalized Eaton/Luneburg lens with $a = 4$ (red) and the related lens from 6-12 potential (green). (c) The ray trajectories from a point of (1, 0) in the generalized lens. (d) The ray trajectories from a point of (1, 0) in the related lens from 6-12 potential.*

In summary, we found the intrinsic refractive index profile for Eaton lens and Luneburg lens, which is from Morse potential. Such a conformal lens has an asymmetric self-focusing effect. We also proposed a generalized Eaton/Luneburg lens and discussed their imaging properties. All the lenses will be very useful in future conformal cloaking designs and other imaging functionalities [5, 7, 14]. In addition, if the lens become finite, it would be possible to design an omnidirectional concave lens, just like the Luneburg lens is an omnidirectional convex lens. Finally, it would be also very promising to find more of this kind of intrinsic potentials and design the related conformal lenses [12].

**Acknowledgements**

This work was supported by the National Science Foundation of China (Grant No. 11874311).